\documentclass[preprint,12pt,preprintnumbers,amsmath,amssymb,floatfix,endfloats*]{revtex4}

\usepackage{graphicx}
\usepackage{dcolumn}
\usepackage{bm}
\usepackage{amsfonts}

\DeclareMathAlphabet{\mathsfsl}{OT1}{cmr}{bx}{it}
\begin{document}
\title{Effects of crossflow velocity and transmembrane pressure on microfiltration of oil-in-water emulsions}
\author{Tohid Darvishzadeh and Nikolai V. Priezjev}
\affiliation{Department of Mechanical Engineering, Michigan State
University, East Lansing, Michigan 48824}
\date{\today}
%
\begin{abstract}

This study addresses the issue of oil removal from water using
hydrophilic porous membranes.    The effective separation of
oil-in-water dispersions involves high flux of water through the
membrane and, at the same time, high rejection rate of the oil
phase. The effects of transmembrane pressure and crossflow velocity
on rejection of oil droplets and thin oil films by pores of
different cross-section are investigated numerically by solving the
Navier-Stokes equation.   We found that in the absence of crossflow,
the critical transmembrane pressure, which is required for the oil
droplet entry into a circular pore of a given surface
hydrophilicity, agrees well with analytical predictions based on the
Young-Laplace equation.    With increasing crossflow velocity, the
shape of the oil droplet is strongly deformed near the pore entrance
and the critical pressure of permeation increases.   We determined
numerically the phase diagram for the droplet rejection, permeation,
and breakup depending of the transmembrane pressure and shear rate.
Finally, an analytical expression for the critical pressure in terms
of geometric parameters of the pore cross-section is validated via
numerical simulations for a continuous oil film on elliptical and
rectangular pores.

\vskip 0.1in



\end{abstract}


\maketitle

\section{Introduction}

With the recent advances in environmental and biological
technologies, there has been increasing interest in characterization
and modeling flows at the micron scales including flows in
microchannels and nanochannels~\cite{Squires05,Eijkel05}, multiphase
flows through porous media~\cite{Blunt01,Coutelieris12}, and
droplet-based microfluidics~\cite{Shui07,Teh08}.    The industrial
applications include oil extraction from porous
media~\cite{McAuliffe73,Kokal92,Sheorey01}, treatment of oily
wastewater~\cite{Mueller97,Cheryana98,Chang01}, and encapsulation of
molecules, cells, and
microorganisms~\cite{Orive03,Atencia05,Zhang12}.      In most of
these processes, one or more phases are dispersed in a continuous
phase in the form of emulsions, which are usually produced by
shearing two immiscible phases against each other in the presence of
surfactants~\cite{Ficheux98}.    In some cases, emulsions serve as
means of transport of molecules, bio-reagents, and drugs, and
ultimately provide the environment for enhanced
reactions~\cite{Atencia05,Kelly07}.    Another technological
application of emulsions is to improve the transportability or
displacement of highly viscous liquids.   For example, heavy crude
oil is emulsified to form a less viscous mixture to facilitate its
transportation~\cite{Langevin04,Sanchez94}.    In addition,
oil-in-water emulsions are used to enhance recovery and increase
sweep efficiency from crude oil reservoirs by blocking highly
permeable paths and preventing channeling of the displacing
fluid~\cite{McAuliffe73}.   Common methods for separation of
emulsions include evaporation of the continuous
phase~\cite{Gryta99}, destruction
(demulsification)~\cite{Bibette99}, and membrane
filtration~\cite{Mueller97}.

Membrane microfiltration has proven to be an efficient way for
separating oil-in-water emulsions~\cite{Pan07,Gryta01}. In
comparison with the conventional methods of filtration (gravity
separators, centrifuges, etc.), membrane microfiltration has several
distinct advantages including reduced space requirements, higher
permeate quality, and lower operating costs~\cite{Mueller97}.
Despite its advantages, microfiltration efficiency can be greatly
reduced because of membrane fouling at highly concentrated emulsions
or long filtration times~\cite{Song98}.    Fouling is generally
caused by the accumulation of the rejected phase on the surface of
the membrane or inside the pore.    There are four main mechanisms
(blocking laws) for membrane fouling, i.e., standard blocking,
complete blocking, intermediate blocking, and cake
formation~\cite{Hermia82}.    Complete blocking is common for very
dilute mixtures and during the initial stages of filtration when
some pores are sealed by droplets and particles, thus reducing the
permeate flux~\cite{Heertjes57,Bowen95}.    Accumulation of the
rejected droplets on the membrane surface results in the formation
of the so-called cake layer, which is sometimes referred to as the
secondary membrane as it adds a hydraulic resistance to the
microfiltration process~\cite{Kuberkar00,Dufreche02,Tarabara04}.
This mechanism is dominant at the final stages of filtration when
the water flux depends mainly on the thickness of the cake layer.

The efficiency of the microfiltration process is determined by the
properties of the membrane material and oil-in-water mixtures.   For
example, the permeate flux is highly dependent on the oil
concentration, stability of the oil phase in water, and the size
distribution of oil droplets~\cite{Mueller97,Park01}.    Moreover,
the membrane properties such as membrane material, pore size and
morphology, and membrane geometry affect the permeate flow
resistance~\cite{Mueller97,Ohya98}.    It was shown that slotted
(rectangular) pores resulted in higher flux rates compared to
circular pores for similar operating conditions because of the lower
fowling rate of the slotted-pore membranes~\cite{Bromley02}. Another
approach to reduce the fouling rate is to introduce crossflow above
the membrane surface.  This method, known as the ``crossflow
microfiltration", reduces fouling by sweeping away the deposited
drops and particles and, hence, decreases the thickness of the cake
layer.      Therefore, crossflow microfiltration systems tend to
produce higher permeate fluxes for longer times compared to dead-end
microfiltration systems~\cite{Koltuniewicz95,Holdich98}. One of the
aims of the present study is to investigate numerically the entry
dynamics of oil droplets into a membrane pore in the presence of
crossflow.

The dynamics of droplet breakup in steady shear flow is determined
by the relative competition of the viscous stress, pressure, and
interfacial tension~\cite{Stone94}.    In general, the breakup
process is initiated by the droplet deformation, which is linearly
proportional to the rate of shear~\cite{Grace82}.    When the
critical deformation is reached, the droplet assumes an unstable
configuration and undergoes a transient elongation before it breaks
up~\cite{Stone94}.     It was also shown that the geometric
confinement as well as the viscosity ratio of the dispersed and
continuous phases influence droplet breakup~\cite{Vananroye06}. In
recent years, the problem of droplet deformation and breakup has
been extensively studied numerically using Lattice
Boltzmann~\cite{Xi99,Graaf06,Fakhari10}, boundary
integral~\cite{Li97,Janssen07}, and Volume of Fluid
(VOF)~\cite{Li00,Dietsche09,Gueyffier99} methods.   The VOF method
used in the present study has proven to be a powerful and efficient
interface tracking algorithm that is both conceptually simple and
relatively accurate~\cite{Gopala08}.    Due to the conservative
discretization of the governing equations in the VOF method, the
mass of each fluid is accurately conserved~\cite{Li00,Pilliod04}.
Furthermore, the ability of the VOF method to automatically capture
local and global changes of the interface topology, e.g.,
coalescence and breakup of droplets, has made it attractive for
various multiphase flow applications~\cite{Gopala08}.

During the last decade, a number of studies have investigated the
process of droplet formation using cross-flowing streams in T-shaped
junctions~\cite{Xu05,Husny06,Graaf06,Gupta09}.     In these
microfluidic systems, two immiscible liquids are driven through
separate channels until their streams meet at a junction, where the
dispersed liquid extends into the continuous stream, resulting in
periodic formation of equal-sized droplets~\cite{Christopher07}.
Regardless of the specific channel geometry and wettability of the
channel walls, breakup of the emerging droplet in a cross-flowing
stream is determined by the viscous drag when the droplet remains
unconfined by the microchannel~\cite{Christopher07}.   For
unconfined T-junctions, it was demonstrated experimentally that the
droplet size strongly depends of the crossflow rate of the
continuous phase and only weakly on the flow rate of the dispersed
phase~\cite{Xu05,Husny06}. It was also shown that, for given value
of continuous phase flow rate, the size of oil droplets decreases
with increasing viscosity ratio of the oil and water~\cite{Husny06}.

In this paper, numerical simulations based on the Volume of Fluid
method are performed to study the influence of transmembrane
pressure and crossflow velocity on the entry dynamics of thin oil
films and droplets into pores of various cross-sections.  We find
that the formula derived in Ref.\,\cite{Nazzal96} for the critical
pressure of permeation of an oil droplet into a circular pore agrees
well with the results of numerical simulations.  The numerical
analysis is then extended to thin oil films covering pores with
elliptical and rectangular cross-section in the absence of
crossflow.  In the presence of crossflow, we obtain numerically the
phase diagram for the droplet rejection, permeation, and breakup as
a function of the transmembrane pressure and shear rate, and study
the details of the processes in three different regions of the phase
diagram.  These results are relevant to microfiltration of dilute
oil-in-water emulsions at early stages before the formation of the
cake layer.

The rest of the paper is organized as follows.   The details of
numerical simulations are described in the next section.   The
analytical predictions based on the Young-Laplace equation are
reviewed in Sec.~\ref{subsection:analytical} and verified
numerically for an oil droplet on a circular pore in Sec.~\ref
{subsection:circular_pores}.   The critical pressure of permeation
for pores with elliptical and rectangular cross-section is reported
in Sec.~\ref {subsection:non-circular}.  The results for the oil
droplet dynamics near circular pores in the presence of crossflow
are presented in Sec.~\ref{subsection:circular_shear}.   The
conclusions are given in the last section.

\section{Details of numerical simulations}
\label{sec:Model}

\begin{figure}[t]
\includegraphics[width=7.0cm,angle=0]{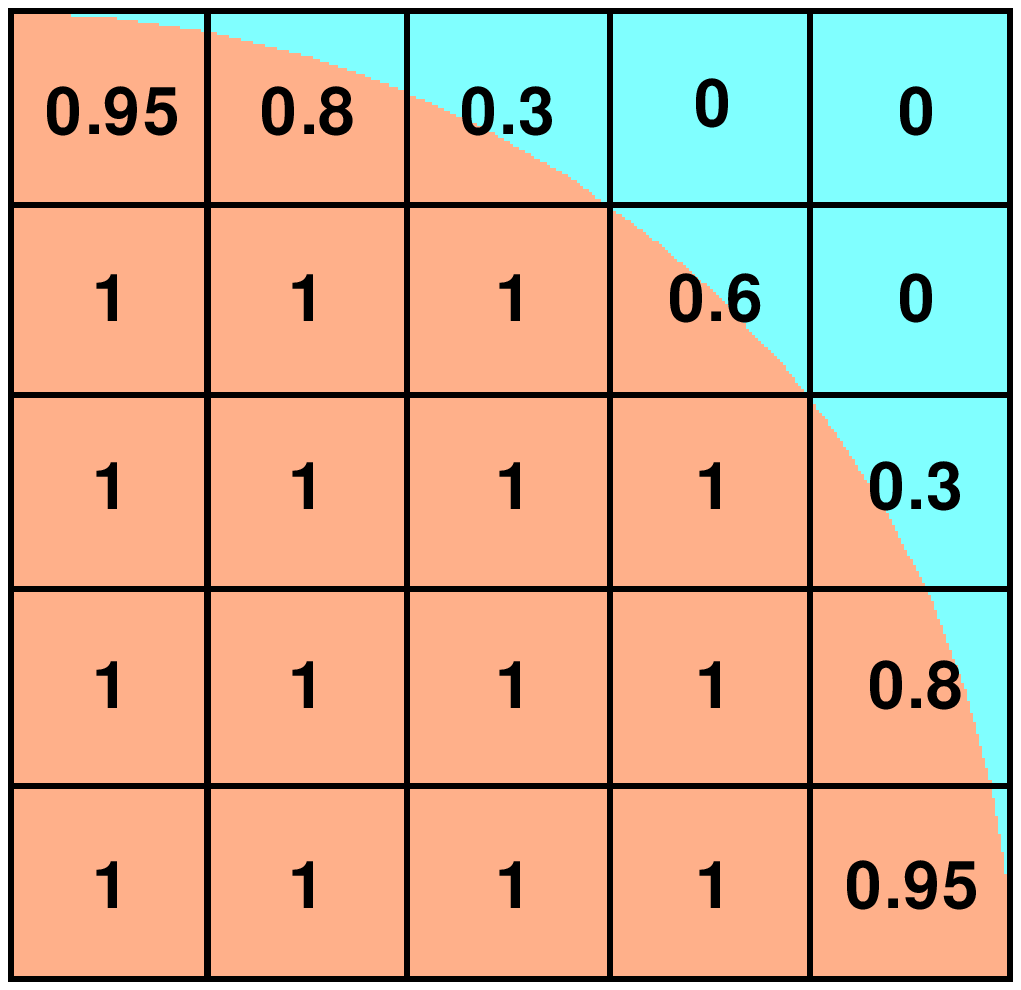}
\caption{(Color online) Values of the volume fraction for each of
the phases and the interface ($\alpha=1$ in the first phase,
$\alpha=0$ in the second phase, and $0<{\alpha}<1$ at the
interface). } \label{fig:vof}
\end{figure}

Numerical simulations were carried out using the commercial software
FLUENT~\cite{fluent}.  In order to control the transmembrane
pressure and the crossflow velocity, a user-defined function was
written and compiled along with the main solver.  The Volume of
Fluid method was used to solve the multiphase flow
problem~\cite{Hirt81}.  For a two-phase fluid, this method is based
on the fact that the two phases form an impenetrable interface,
i.e., each cell is filled with either one of the phases (denoting a
specific phase zone) or a combination of two phases (denoting the
interface).  This is achieved by introducing a variable $\alpha$,
known as the ``volume fraction", which is defined as the ratio of
the volume of fluid in the cell and the total cell volume; and it
varies between $0$ and $1$~\cite{Saha09,Hsu04}.  An example of how
the volume fraction varies near the interface is illustrated
schematically in Fig.\,\ref{fig:vof}.


The interface is tracked by solving the transport equation for the
volume fraction as follows:
\begin{equation}
\frac{\partial{\alpha}}{{\partial}{t}}+{\nabla}{\cdot}\,({\alpha}{\textbf{V}})=\,0,
\label{eq:continuity}
\end{equation}
where $\textbf{V}$ is the velocity vector.
Equation\,(\ref{eq:continuity}) states that the substantial
derivative of the volume fraction is zero, and, therefore, the
interface is convected by the velocity fields at the interface.
After solving Eq.\,(\ref{eq:continuity}), the material properties
are computed by considering the fraction of each component in the
cell; e.g., the density is given by
\begin{equation}
\rho = \alpha\,\rho_{2}+(1-\alpha)\,\rho_{1},
\label{eq:compatibility}
\end{equation}
where $\rho$ is the volume-fraction-averaged density.

One momentum equation is solved and the velocity field is shared
between two phases as follows:
\begin{equation}
\frac{\partial}{{\partial}{t}}({\rho}{\textbf{V}})+\nabla\cdot(\rho{\textbf{V}}{\textbf{V}})
=-\nabla{p}+\nabla\cdot[\mu(\nabla{\textbf{V}}+\nabla{\textbf{V}}^{T})]+\rho\,\textbf{g}+\textbf{F},
\label{eq:momentum}
\end{equation}
where $\textbf{g}$ is the vector of gravitational acceleration, and
$\textbf{F}$ is the source term.   In multiphase flow applications,
the source term is the surface tension force per unit volume and it
is non-zero only at the interface.   Using the divergence theorem,
the surface tension force is defined as the volume force in a cell
as follows:
\begin{equation}
F_{\sigma} =
\sigma\,\frac{\rho\,\kappa\nabla\alpha}{\frac{1}{2}(\rho_{1}+\rho_{2})},
\label{eq:surface_tension}
\end{equation}
where $\sigma$ is the surface tension between two phases and
$\kappa$ is the mean curvature of the interface in the cell.  This
force is related to the pressure jump across the interface
(determined by the Young-Laplace equation) and it acts in the
direction normal to the interface.  The surface tension term tends
to smooth out regions with large interface
curvature~\cite{Gerlach06}.  If the interface is in contact with the
wall, the normal vector ($\nabla\alpha$), which defines the
orientation of the interface in the cell adjacent to the wall, is
determined by the contact angle.   The effect of static contact
angle is taken into account by imposing the interface unit normal
for a point (a cell in the Finite Volume method), $\textbf{n}_{i}$,
on the wall containing the interface as follows:
\begin{equation}
\textbf{n}_{i}=\,\textbf{n}_{w}\,\textrm{cos}(\theta_{st})+\,\textbf{n}_{t}\,\textrm{sin}(\theta_{st}),
\label{eq:contact_angle}
\end{equation}
where $\textbf{n}_{w}$ is the unit vector normal to the wall,
$\textbf{n}_{t}$ is a vector on the wall and normal to the contact
line, and $\theta_{st}$ is the static contact
angle~\cite{Brackbill92}.

A SIMPLE algorithm was used for the pressure-velocity decoupling.
The momentum equation was discretized using a second order upwind
scheme.  To reconstruct the interface and, consequently, solve the
volume fraction transport equation, a PLIC (Piecewise Linear
Interface Reconstruction) method was used~\cite{Rider98}.   The
pressure equation is discretized using a staggered mesh with central
differencing.  In FLUENT, the interfacial tension is modeled using
the well-known model of Continuum Surface Force (CSF) of Brackbill
\textit{et al.}~\cite{Brackbill92}.   Using CSF, the surface tension
volume force Eq.\,(\ref{eq:surface_tension}) is added as a source
term to the momentum equation and the curvature is computed in terms
of the vector normal to the interface $\textbf{n}$ via:
\begin{equation}
\kappa=\,\frac{1}{|\textbf{n}|}\Big[\Big(\frac{\textbf{n}}{|\textbf{n}|}\cdot\,\nabla\Big)|\textbf{n}|-
(\nabla\cdot\,\textbf{n})\Big].
\label{eq:curvature}
\end{equation}

Interfacial effects in multiphase flows are described by the
Young-Laplace equation, which relates the pressure jump across the
interface to its mean curvature and the surface tension coefficient.
For flows at the micron length scales, the viscous effects are
dominant and the inertial effects are typically negligible.   The
capillary number is a measure of how viscous shear stresses are
compared to the interfacial stresses and it is defined
$Ca={\mu}\,U/{\sigma}$, where $U$ is the characteristic velocity,
${\mu}$ is the fluid viscosity, and ${\sigma}$ is the surface
tension coefficient.

In the present study, the numerical simulations are performed to
investigate a separation process of two commonly used liquids, i.e.,
kerosene and water.   The density of kerosene is
$\rho_{o}=889\,\text{kg/m}^3$ and the viscosity ratio of kerosene
and water at standard conditions is $\mu_{o}/\mu_{w}=2.4$.   It is
assumed that the water is deionized; and, thus, the surface tension
coefficient $\sigma=0.0191\,\text{N/m}$ is used throughout the
study~\cite{Cumming00}.  Furthermore, we consider hydrophilic
surfaces (for example polyvinyl-pyrrolidone~\cite{Cumming00}) with
contact angles of kerosene in water greater than $90^{\circ}$.

In our simulations, the mesh was generated in GAMBIT using the
Cooper mesh scheme.  This method works by sweeping the node patterns
of specified source faces through the whole volume and the resultant
mesh consists of an array of tetrahedral grids.  For the results
reported in the current study, we used about $30$ cells along the
pore diameter.    To test the grid-resolution dependence, we
considered $3$ finer meshes that contained $50$, $70$, and $90$
cells along the pore diameter.    In the absence of crossflow, the
simulations were performed for an oil droplet
($r_d=1.0\,{\mu}\text{m}$) on a circular pore
($r_p=0.2\,{\mu}\text{m}$) at two transmembrane pressures ($1.000$
and $0.951$ bar) slightly above and below the exact value of the
permeation pressure $0.976$ bar predicted by the Young-Laplace
analysis.   In all cases, the droplet would either penetrate into
the pore or reside at the pore entrance for at least
$40\,{\mu}\text{s}$.    Furthermore, it was previously shown that
the velocity of the contact line in the VOF method is inversely
proportional to the logarithm of the mesh size~\cite{Renardy01},
and, therefore, it is expected that the droplet velocity in the
shear flow will depend on the grid resolution.   However, in the
present study, the oil droplet becomes temporarily pinned at the
pore entrance by the transmembrane pressure, and thus the contact
line velocity becomes much smaller than the flow velocity in the
channel.   Nevertheless, we have performed numerical simulations in
the permeation, rejection, and breakup regions of the phase diagram
with $4$ times finer meshes and found that our results remain
unchanged.


\section{Results}
\label{sec:Results}

\subsection{The Young-Laplace analysis for circular pores}
\label{subsection:analytical}

Effective separation of oil-in-water emulsions is controlled by
several key parameters such as the membrane pore size, surface
energy, size of oil droplets, surface tension, and pressure
difference across the membrane.   If the transmembrane pressure is
relatively high, then oil droplets will most likely penetrate the
membrane surface resulting in low rejection rates of the oil phase.
On the other hand, low transmembrane pressures tend to limit flux of
water through the membrane.   Hence, the optimum operating
conditions strongly depend on the critical transmembrane pressure
required for an oil droplet entry into a membrane pore.

When the transmembrane pressure across a hydrophilic membrane
exceeds a certain critical value, the oil phase will penetrate the
membrane.   Thus, for high separation efficiency, the transmembrane
pressure should be maintained at a value below $P_{crit}$, which for
a continuous oil film on the membrane surface with circular pores is
given by
\begin{equation}
P_{crit} = \frac{2\,\sigma\,\textrm{cos}\,\theta}{r_{p}},
\label{eq:pcrit}
\end{equation}
where $\sigma$ is the surface tension coefficient between oil and
water, $\theta$ is the contact angle of the interface of an oil
droplet on a membrane surface immersed in water, and $r_{p}$ is the
membrane pore radius~\cite{Nazzal96}.   The critical pressure in
Eq.\,(\ref{eq:pcrit}) is determined by the Young-Laplace pressure
due to the curvature of the oil-water interface inside the pore.

If instead of a thin oil film, a droplet of oil is placed at the
entrance of the membrane pore, the formula for the critical
pressure, Eq.\,(\ref{eq:pcrit}), has to be corrected by a factor
that depends on the ratio $r_{d}/r_{p}$ to include the effect of the
oil-water interface curvature above the membrane surface.  It was
previously shown~\cite{Cumming00,Nazzal96} that the pressure
required to force an entry of an oil droplet of radius $r_{d}$ into
a circular pore is given by
\begin{equation}
P_{crit} =
\frac{2\,\sigma\,\textrm{cos}\,\theta}{r_{p}}\,\,\sqrt[3]{1-\frac{2+3\,\textrm{cos}\,\theta-
\textrm{cos}^3\theta}{4\,(r_{d}/r_{p})^3\,\textrm{cos}^3\theta-(2-3\,\textrm{sin}\theta+\textrm{sin}^3\theta)}}.
\label{eq:pcritfull}
\end{equation}
In the limit $r_{d}\rightarrow\infty$, the curvature of the droplet
above the pore vanishes, and thus this formula corresponds to a
continuous oil film on the membrane surface, and
Eq.\,(\ref{eq:pcritfull}) converges to Eq.\,(\ref{eq:pcrit}).
Contrary to the case of the thin film, the critical pressure for an
oil droplet with $\theta=90^{\circ}$ is negative, and the droplet
will penetrate into the pore in the absence of the applied pressure
gradient.   We also comment that no such expression exists for the
case when crossflow is present and the hydrodynamic drag force is
exerted on the droplet parallel to the membrane surface.

In what follows, we investigate the dynamics of an oil droplet and
thin film entry into pores with various cross-sections.    The
critical pressures [Eq.\,(\ref{eq:pcrit}) and
Eq.\,(\ref{eq:pcritfull})] are compared with the results of
numerical simulations in order to validate the numerical scheme. The
numerical analysis is then extended to pores with rectangular and
elliptical cross-sections.   Finally, the effect of shear flow on
the droplet entry is considered and a phase diagram of the
transmembrane pressure versus shear rate is determined numerically.

\subsection{An oil droplet on a circular pore in the absence of crossflow}
\label{subsection:circular_pores}

We first present the numerical results for the critical pressure
required to force an entry of an oil droplet into a cylindrical
pore.   In the numerical scheme, the oil droplet is initially
immersed in water above the membrane surface in the absence of flow.
The transmembrane pressure is then gradually increased to a
predetermined value.    As the simulation continues, the droplet
approaches the membrane surface and resides at the pore entrance.
Depending on the applied pressure difference across the membrane,
the droplet will either remain at the pore entrance or penetrate
into the pore.   We comment that when the applied pressure is close
to the critical pressure, the dynamics of an oil droplet entry into
the pore is significantly slowed down because the net driving force
on the droplet is reduced.

\begin{figure}[t]
\includegraphics[width=13.0cm,angle=0]{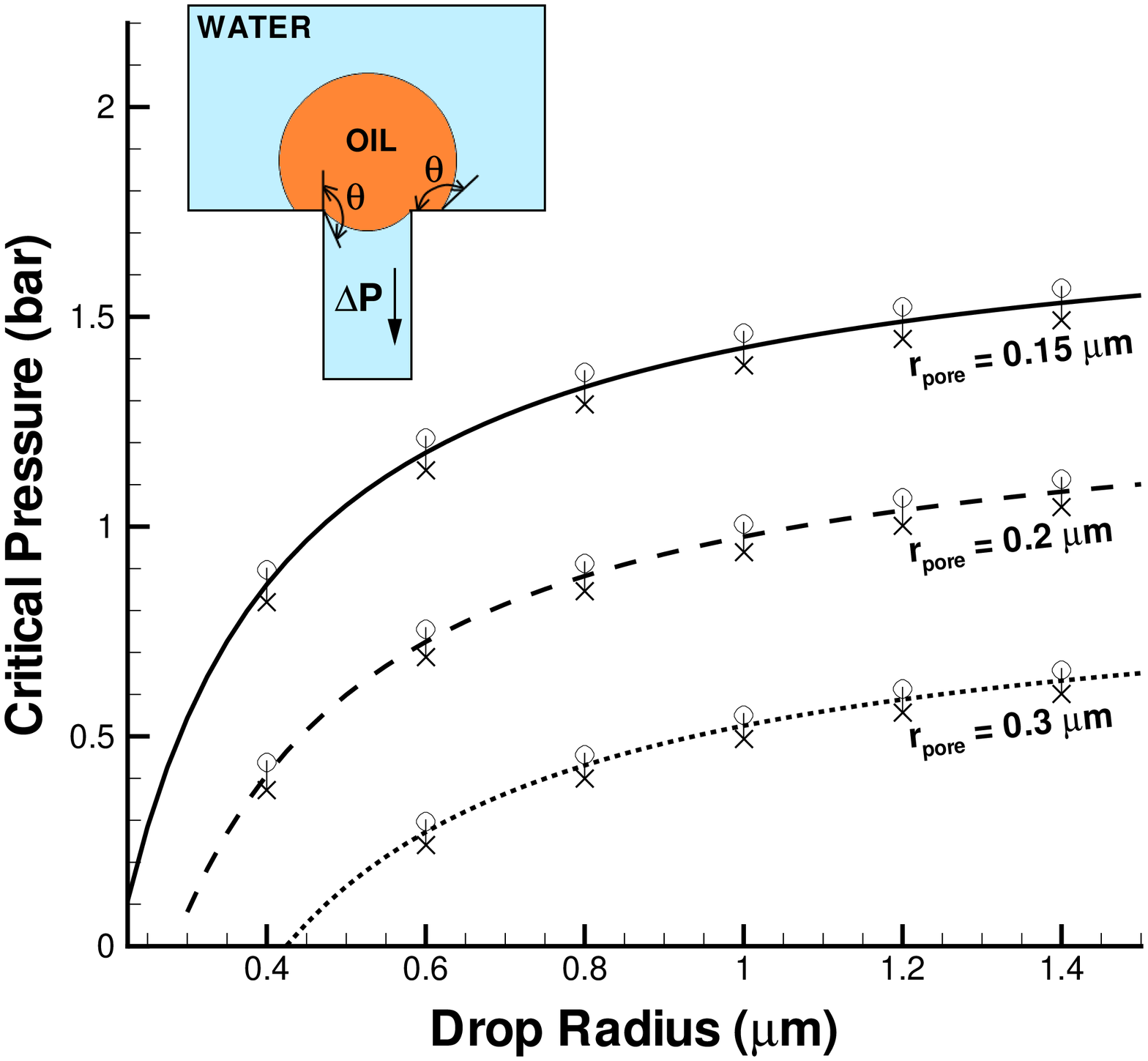}
\caption{(Color online)  The critical pressure computed using
Eq.\,(\ref{eq:pcritfull}) for the values of the pore radius
$0.15\,{\mu}\text{m}$ (continuous line), $0.2\,{\mu}\text{m}$
(dashed line), and $0.3\,{\mu}\text{m}$ (dotted line).   Error bars
are extracted from the numerical simulations (see text for details)
with the parameters $\mu_{o}/\mu_{w}=2.4$,
$\rho_{o}/\rho_{w}=0.781$, $\sigma=0.0191\,\text{N/m}$, and
$\theta=\,135^{\circ}$.   The symbols indicate ($\times$) rejection
and ($\circ$) permeation of the oil droplet. }
\label{fig:pcrit_rd_rp}
\end{figure}

The critical pressure as a function of the droplet radius is plotted
in Fig.\,\ref{fig:pcrit_rd_rp} using Eq.\,(\ref{eq:pcritfull}) for
three values of the pore radius.   The error bars in
Fig.\,\ref{fig:pcrit_rd_rp} indicate the upper and lower values of
the transmembrane pressure when the oil droplet either enters the
pore or remains at the pore entrance during the time interval of
about $40\,{\mu}\text{s}$.  We find an excellent agreement between
the results of numerical simulations and analytical predictions of
Eq.\,(\ref{eq:pcritfull}), which provides a validation of the
numerical method.   Furthermore, as shown in
Fig.\,\ref{fig:pcrit_rd_rp}, the critical pressure increases with
increasing droplet radius or decreasing pore radius.  As the droplet
radius increases, the critical pressure approaches an asymptotic
value predicted by Eq.\,(\ref{eq:pcrit}) for a thin oil film (not
shown).

\subsection{Critical pressure for pores with arbitrary cross-section}
\label{subsection:non-circular}

We next investigate the influence of pore cross-sectional shape on
the critical transmembrane pressure using simple physical arguments
and numerical simulations.   According to the Young-Laplace
equation, the pressure jump across an interface between two
immiscible fluids is related to its mean curvature and the surface
tension as follows:
\begin{equation}
{\Delta}P = 2\,{\sigma}\,\kappa,
\label{eq:young}
\end{equation}
where $\kappa$ is the mean curvature of the interface computed by
averaging two principle curvatures.  In the absence of gravity, the
mean curvature of an arbitrary surface $z(x,y)$ is given by
\begin{equation}
2\,\kappa =
\nabla\cdot\Big(\frac{{\nabla}\,z}{\sqrt{1+|{\nabla}\,z|^{2}}}\Big).
\label{eq:laplace}
\end{equation}
It follows from Eq.\,(\ref{eq:young}) that an interface, which is
subject to a prescribed pressure jump and constant surface tension
coefficient, has a constant mean curvature.   Therefore, if the
gravity is negligible, the fluid-fluid interface forms a section of
the so-called Delaunay surface~\cite{Hill11,Eells87}.

\begin{figure}[t]
\includegraphics[width=8.0cm,angle=0]{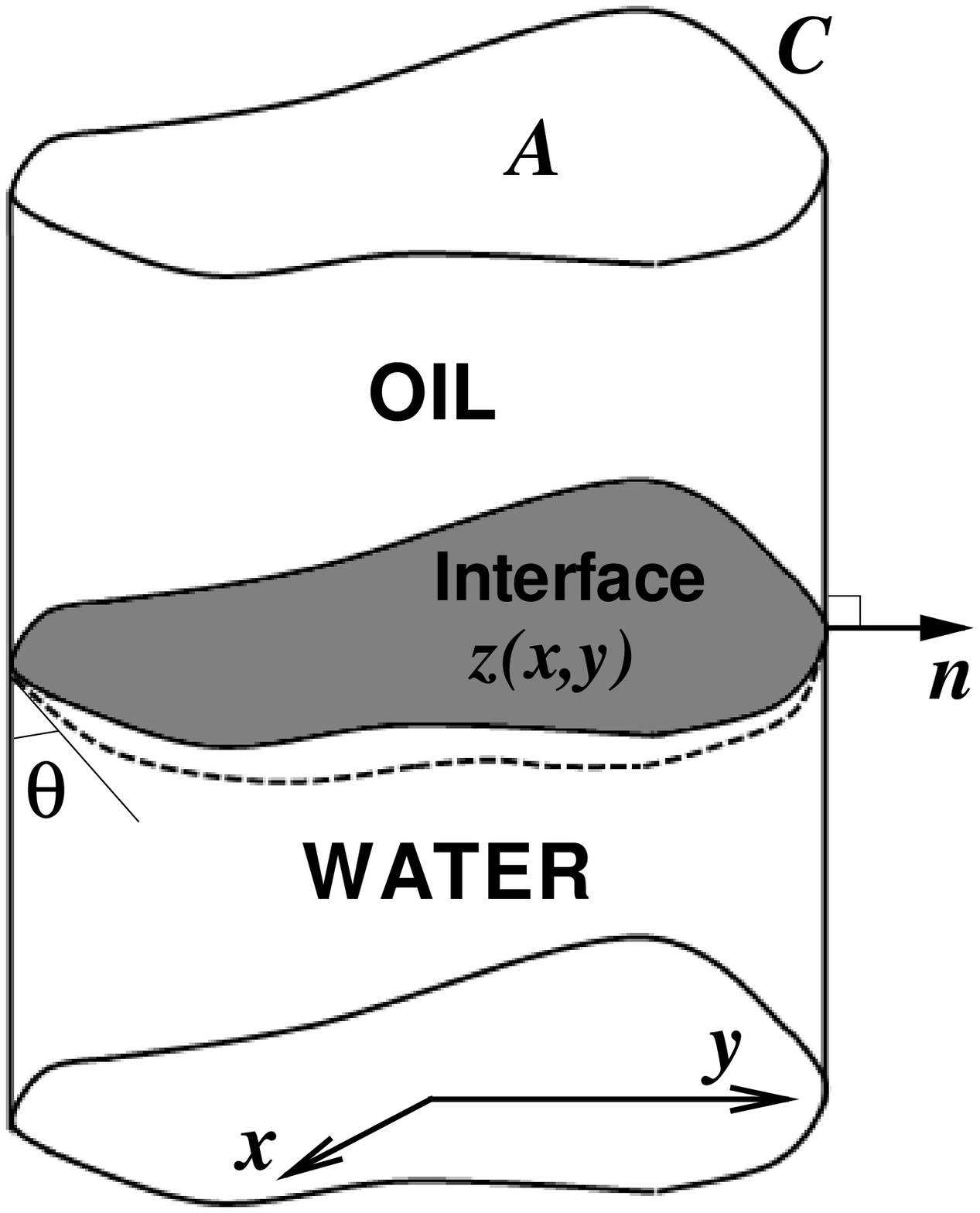}
\caption{Schematic representation of the oil-water interface inside
the pore of arbitrary cross-section with perimeter $C$ and area $A$.
The interface forms a constant angle $\theta$ with the inner surface
of the pore. } \label{fig:arb_schem}
\end{figure}

As shown in Fig.\,\ref{fig:arb_schem}, if the oil-water interface is
bounded by the walls of a pore of arbitrary cross-section, the
constant contact angle at the pore surface imposes a boundary
condition for Eq.\,(\ref{eq:laplace}) in the form
\begin{equation}
\textrm{cos}\,\theta =
\textbf{n}\cdot\Big(\frac{{\nabla}\,z}{\sqrt{1+|{\nabla}\,z|^{2}}}\Big),
\label{eq:contact_angle}
\end{equation}
where \textbf{n} is the outward unit vector normal to the pore
surface~\cite{Finn99}.

Integrating Eq.\,(\ref{eq:laplace}) over an arbitrary cross-section
with a smooth boundary and using the divergence theorem, we obtain
\begin{equation}
A_{p} \cdot 2\,\kappa
=\int_{C_{p}}\textbf{n}\cdot\Big(\frac{{\nabla}\,z}{\sqrt{1+|{\nabla}\,z|^{2}}}\Big)\,dL,
\label{eq:young_int_div}
\end{equation}
where $C_{p}$ and $A_{p}$ are the perimeter and cross-sectional area
respectively.  Taking the integral on the right hand side over the
perimeter gives the following relation for the mean curvature of the
oil-water interface~\cite{Concus90}
\begin{equation}
2\,\kappa =\frac{C_{p}\,\textrm{cos}\,{\theta}}{A_{p}}.
\label{eq:mean_curv}
\end{equation}
According to Eq.\,(\ref{eq:mean_curv}), the mean curvature of an
interface bounded by a pore of arbitrary cross-section can be
related to the geometric properties of the boundary.  Remember that
Eq.\,(\ref{eq:young}) relates the mean curvature to the surface
tension and the pressure jump.  In the case of an oil film on a pore
with an arbitrary cross-section, the critical applied pressure is
equal to the pressure jump at the interface.    Therefore, combining
Eqs.\,(\ref{eq:young}) and (\ref{eq:mean_curv}), we obtain the
critical permeation pressure for the oil film to enter into a pore
of arbitrary cross-section
\begin{equation}
P_{crit} = \frac{\sigma\,C_p\,\textrm{cos}\,\theta}{A_p}.
\label{eq:force_balance}
\end{equation}
This equation can also be derived from the force balance between the
applied pressure and the Laplace pressure due to the curvature of
the oil-water interface inside the pore.

It should be noted that the boundary value problem
Eq.\,(\ref{eq:laplace}) does not always have a solution for a stable
interface with a constant mean curvature and a constant contact
angle~\cite{Concus74}.  In other words, there is a limitation on the
values of the contact angle that correspond to the attached
interface with a constant mean curvature.   For example, if the
contact angle (computed from the oil phase) is larger than the
critical value, then Eq.\,(\ref{eq:laplace}) subject to the boundary
condition Eq.\,(\ref{eq:contact_angle}) does not have a stable
solution.    As a result, the interface cannot remain attached to
the bounding surface with a prescribed contact angle and, at the
same time, maintain a constant mean curvature required by the
Laplace equation.   The critical contact angle is determined by the
largest curvature of the cross-sectional shape for smooth boundaries
and by the smallest wedge angle for boundaries with sharp
corners~\cite{Brady04}.

In what follows, we consider two special cases of rectangular and
elliptical pores and compare predictions of
Eq.\,(\ref{eq:force_balance}) with the results of numerical
simulations.   The problem is illustrated schematically in the inset
of Fig.\,\ref{fig:slotted_film}.  The oil film covers the pore
entrance of a hydrophilic membrane subject to a pressure gradient
(the transmembrane pressure).   Since the Bond number is small,
$Bo=(\rho_{w}-\,\rho_{o})\,r_{d}^{2}\,g/\,\sigma=6\times10^{-8}$,
the effect of the gravitational force can be neglected.


\subsubsection{Thin oil film on the membrane surface with a rectangular pore}
\label{subsection:slotted_pores}

In this subsection, we investigate the dynamics of an oil film entry
into a rectangular pore, which is sometimes referred to as a
``slotted pore''~\cite{Ullah12}.   The rectangular shape of a
slotted pore is characterized by the width (the shorter side) and
the length (the longer side).  Using $A_{p}=w\,l$ and
$C_{p}=2\,(w+l)$, the corresponding critical pressure is obtained
from Eq.\,(\ref{eq:force_balance}) as follows:
\begin{equation}
P_{crit}=\,2\,\sigma\,\textrm{cos}\,\theta\,\Big(\frac{1}{w}+\frac{1}{l}\Big),
\label{eq:pcr_slotted}
\end{equation}
where $w$ and $l$ are the width and length of the pore
cross-section~\cite{Concus90,Ichikawa04,Cho07}.  In the limit when
$l\,\,{\gg}\,\,w$, Eq.\,(\ref{eq:pcr_slotted}) reduces to $P_{crit}=
2\,\sigma\,\textrm{cos}\,\theta\,/\,w$, which is the critical
pressure on an oil film entering into an infinitely long rectangular
pore.   In this case, one of the curvatures of the interface is zero
and the other curvature is proportional to the width of the pore,
and thus the shape of the interface is a part of a cylinder with the
radius $w/2\,\textrm{cos}\,\theta$.

\begin{figure}[t]
\includegraphics[width=15.0cm,angle=0]{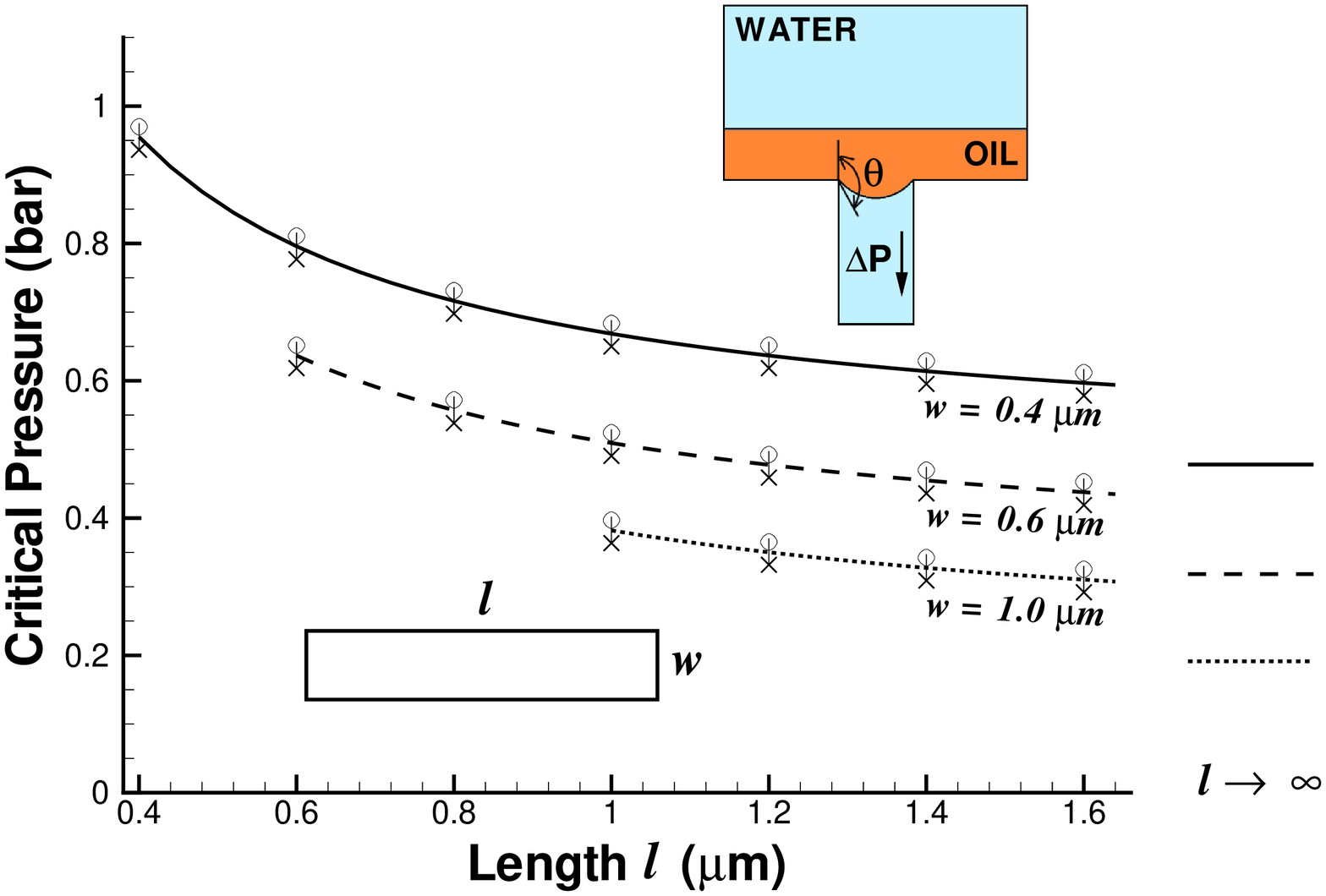}
\caption{ The critical permeation pressure for the oil film into the
rectangular pore with different aspect ratios.  The curves are
Eq.\,(\ref{eq:pcr_slotted}) and the symbols are the numerical
results for $\mu_{o}/\mu_{w}=2.4$, $\rho_{o}/\rho_{w}=0.781$,
$\sigma=0.0191\,\text{N/m}$, and $\theta=\,120^{\circ}$.  The
symbols indicate ($\times$) rejection and ($\circ$) permeation of
the oil film. } \label{fig:slotted_film}
\end{figure}

The results of numerical simulations and predictions of
Eq.\,(\ref{eq:pcr_slotted}) are shown in
Fig.\,\ref{fig:slotted_film} for several aspect ratios.  As
expected, square pores have the highest critical pressure due to the
largest perimeter-to-area ratio.  The critical pressure decreases
with increasing aspect ratio.  These results demonstrate that there
is an excellent agreement between the numerical results and
analytical predictions based on the Young-Laplace equation.  It is
important to note that when the applied pressure is close to the
critical pressure, the net force on the interface is small, and,
therefore, very long simulation time is required to capture the
motion of the interface.  The symbols in
Fig.\,\ref{fig:slotted_film} indicate the rejection and permeation
pressures that were resolved numerically without excessive
computational effort.  Interestingly, each curve in
Fig.\,\ref{fig:slotted_film} is well described by the function
$P_{crit}=2\,\sigma\,\textrm{cos}\,\theta\,/\,l$, which is shifted
upward by a constant $2\,\sigma\,\textrm{cos}\,\theta\,/\,w$
(indicated by the horizontal lines in Fig.\,\ref{fig:slotted_film}).

It was previously shown that for a rectangular cross-section of
arbitrary aspect ratio, Eq.\,(\ref{eq:laplace}) has a solution with
a constant curvature for a non-wetting fluid
($\theta\,>\,90^{\circ}$) when the the contact angle $\theta\leq
135^{\circ}$~\cite{Wong92,Brady04,Ajaev06}.  In the case of a square
pore, the interface is part of a sphere with the radius
$w/2\,\textrm{cos}\,\theta$~\cite{Feng11,Finn99}.  For other aspect
ratios, the interface surface has a constant mean curvature
$\kappa\,=\,\textrm{cos}\,\theta\,(1/w+1/l)$, but it is no longer
spherical~\cite{Concus90}.

\begin{figure}[t]
\includegraphics[width=12.0cm,angle=0]{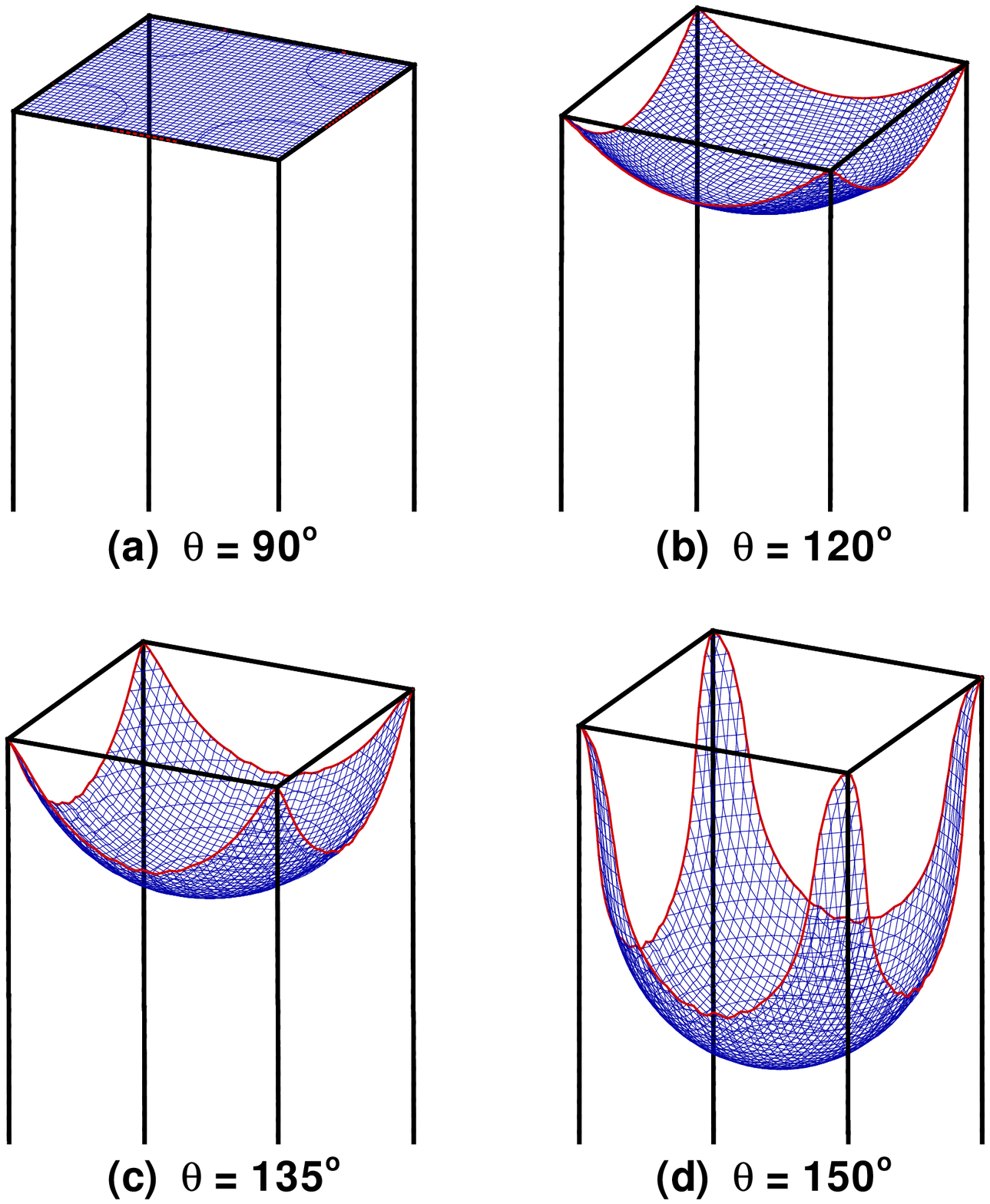}
\caption{ (Color online) Snapshots of the oil-water interface inside
the square pore for contact angles
$\theta=90^{\circ},~120^{\circ},~135^{\circ}$, and $150^{\circ}$.
The critical pressure Eq.\,(\ref{eq:pcr_slotted}) is computed for
the surface tension $\sigma=0.0191\,\text{N/m}$ and the pore width
$1.5\,{\mu}\text{m}$.   Cases (a), (b), and (c) correspond to the
stationary interface, while in the case (d) the interface is in
transient state (see text for details).  } \label{fig:curves}
\end{figure}

Figure\,\ref{fig:curves} shows snapshots of the oil-water interface
inside the square pore obtained from our numerical simulations.  The
transmembrane pressure is set to a value computed from
Eq.\,(\ref{eq:pcr_slotted}) for the contact angles
$\theta=90^{\circ},~120^{\circ},~135^{\circ}$, and $150^{\circ}$. As
observed in Fig.\,\ref{fig:curves}, the concave shape of the
interface strongly depends on the contact angle.  When
$\theta=90^{\circ}$, the interface enters the pore with zero
curvature and, according to Eq.\,(\ref{eq:pcr_slotted}), with zero
pressure gradient.  For contact angles between $90^{\circ}$ and
$135^{\circ}$, the interface bends in the center and penetrates into
the pore before it starts to move near the corners. For
$\theta>135^{\circ}$, the distance between the interface location in
the center and at the corners will theoretically be infinity because
the interface becomes pinned at the corners while the inner part
penetrates into the pore~\cite{Collicott04}.

\subsubsection{Thin oil film on the membrane surface with an elliptical pore}
\label{subsection:elliptical_pores}

The critical permeation pressure for an oil film covering a pore
with an elliptical cross-section can be estimated from
Eq.\,(\ref{eq:force_balance}) and the geometric properties of an
ellipse.  However, there is no exact expression for the perimeter of
an ellipse. In our study, we use one of the most accurate and
compact approximations that predicts the perimeter of an ellipse
with an error of $-0.04\%$~\cite{Berndt85}
\begin{equation}
C_{p}\approx\pi\,(a+b)\Big[1+\frac{3\,h}{10+\sqrt{4-3\,h}}\Big],
\label{eq:pcr_ellipse_perimeter}
\end{equation}
where $a$ and $b$ are the major and minor radii of the ellipse and
$h = (a-b)^{2}/(a+b)^{2}$.  Using
Eq.\,(\ref{eq:pcr_ellipse_perimeter}) and the expression for the
ellipse area $A_{p} = \pi\,a\,b$, the critical pressure is given by
\begin{equation}
P_{crit}\approx\frac{(a+b)}{a\,b}\Big[1+\frac{3\,h}{10+\sqrt{4-3\,h}}\Big]\,\sigma\,\textrm{cos}\,\theta.
\label{eq:pcr_elliptical}
\end{equation}
Clearly, in the case of a circular pore, $a=b=r_{p}$,
Eq.\,(\ref{eq:pcr_elliptical}) is reduced to Eq.\,(\ref{eq:pcrit}).

\begin{figure}[t]
\includegraphics[width=15.0cm,angle=0]{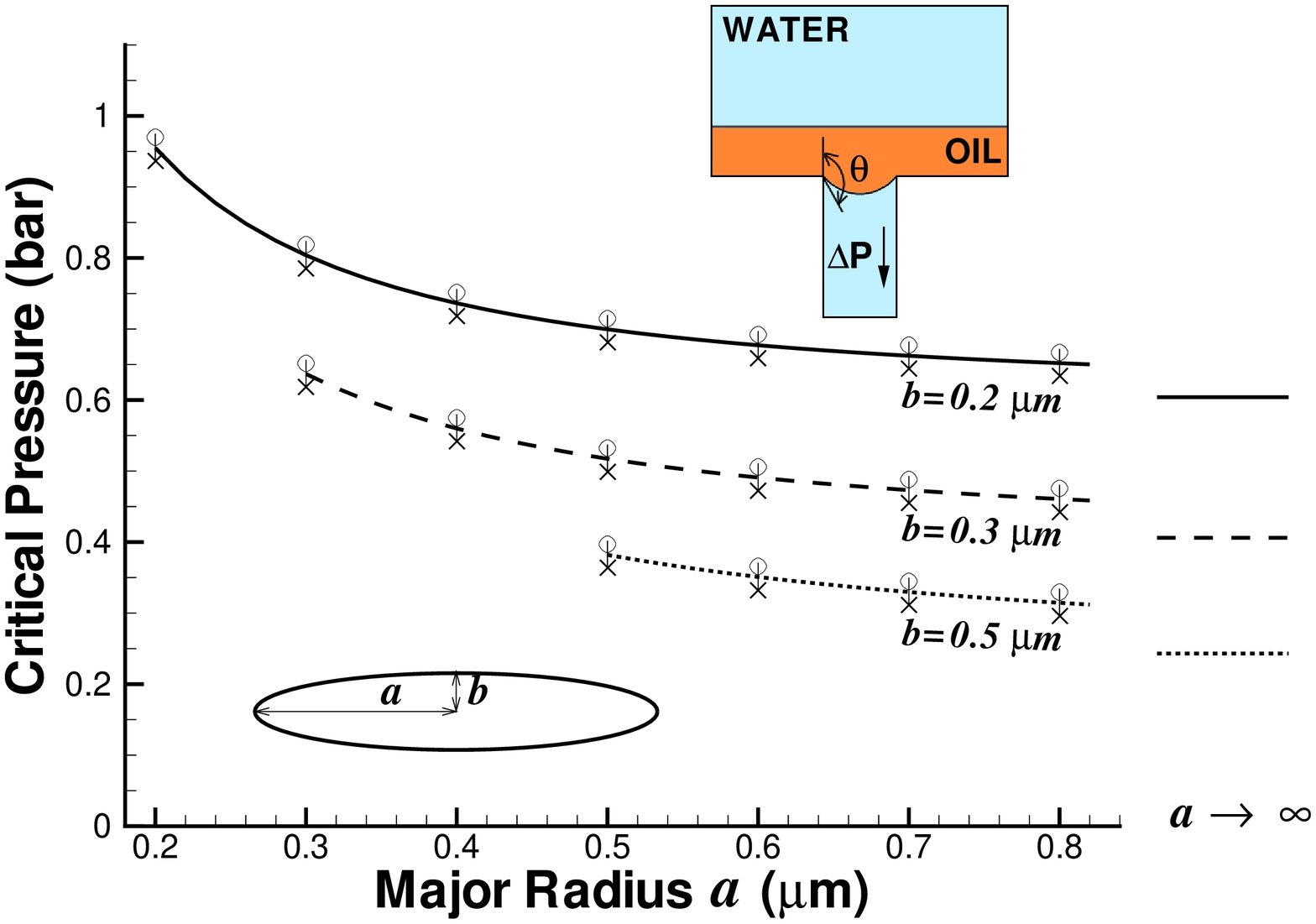}
\caption{ The critical pressure of permeation for the oil film into
the elliptical pore as a function of the major and minor radii.  The
curves are computed using Eq.\,(\ref{eq:pcr_elliptical}).  The
symbols represent numerical results for the parameters
$\mu_{o}/\mu_{w}=2.4$, $\rho_{o}/\rho_{w}=0.781$,
$\sigma=0.0191\,\text{N/m}$, and $\theta=120^{\circ}$.  The symbols
indicate ($\times$) rejection and ($\circ$) permeation of the oil
film. } \label{fig:pcr_elliptical}
\end{figure}

The results of numerical simulations and predictions of
Eq.\,(\ref{eq:pcr_elliptical}) are summarized in
Fig.\,\ref{fig:pcr_elliptical} for different aspect ratios.  Similar
to the case of the rectangular pore, the symbols indicate pressures
of rejection and permeation that were resolved during the simulation
time interval of about $40\,{\mu}s$.  It can be observed that the
critical pressure decreases with increasing ellipse aspect ratio,
and the numerical results agree very well with predictions of
Eq.\,(\ref{eq:pcr_elliptical}).  We also comment that when $a\gg b$,
each curve in Fig.\,\ref{fig:pcr_elliptical} asymptotes to
$P_{crit}\approx 4\,\sigma\,\textrm{cos}\,\theta/\pi\,b$, which is
higher than the value
$P_{crit}=2\,\sigma\,\textrm{cos}\,\theta\,/\,w$ estimated for an
infinitely long rectangular pore (see
section~\ref{subsection:slotted_pores}).  This difference arises
because an infinitely long ellipse and an infinitely long rectangle
of equal width have the same perimeter but different areas.

If the aspect ratio of an elliptical pore is less than $1.635$, then
the boundary value problem given by Eq.\,(\ref{eq:laplace}) has a
solution for any contact angle~\cite{Albright77}.   However, when
$a/b>1.635$, there is a critical contact angle above which
Eq.\,(\ref{eq:laplace}) has no solution~\cite{Albright77}. For
$a/b=1.635$,  the critical contact angle is $180^{\circ}$, and it
decreases with increasing aspect ratio.   The largest aspect ratio
considered in the present study is $a/b=4.0$ for which the critical
contact angle is $153.05^{\circ}$~\cite{Albright77}.   Therefore,
the contact angle of $120^{\circ}$ used in our simulations generated
an interface with a constant mean curvature described by
Eq.\,(\ref{eq:mean_curv}).    We comment that the oil-water
interface inside the elliptical pore is not spherical because the
boundary condition Eq.\,(\ref{eq:contact_angle}) is not satisfied at
the intersection of a sphere and a cylinder with an elliptical
cross-section.    Similar to the rectangular cross-section, we found
that at the critical pressure given by
Eq.\,(\ref{eq:pcr_elliptical}), the contact line is pinned at the
antipodal points of the highest curvature of the ellipse,
$(x,y)=({\pm}a,0)$, while the rest of the interface penetrates into
the pore.

\subsection{Sheared droplet on the membrane surface with a circular pore}
\label{subsection:circular_shear}

\begin{figure}[t]
\includegraphics[width=10.0cm,angle=0]{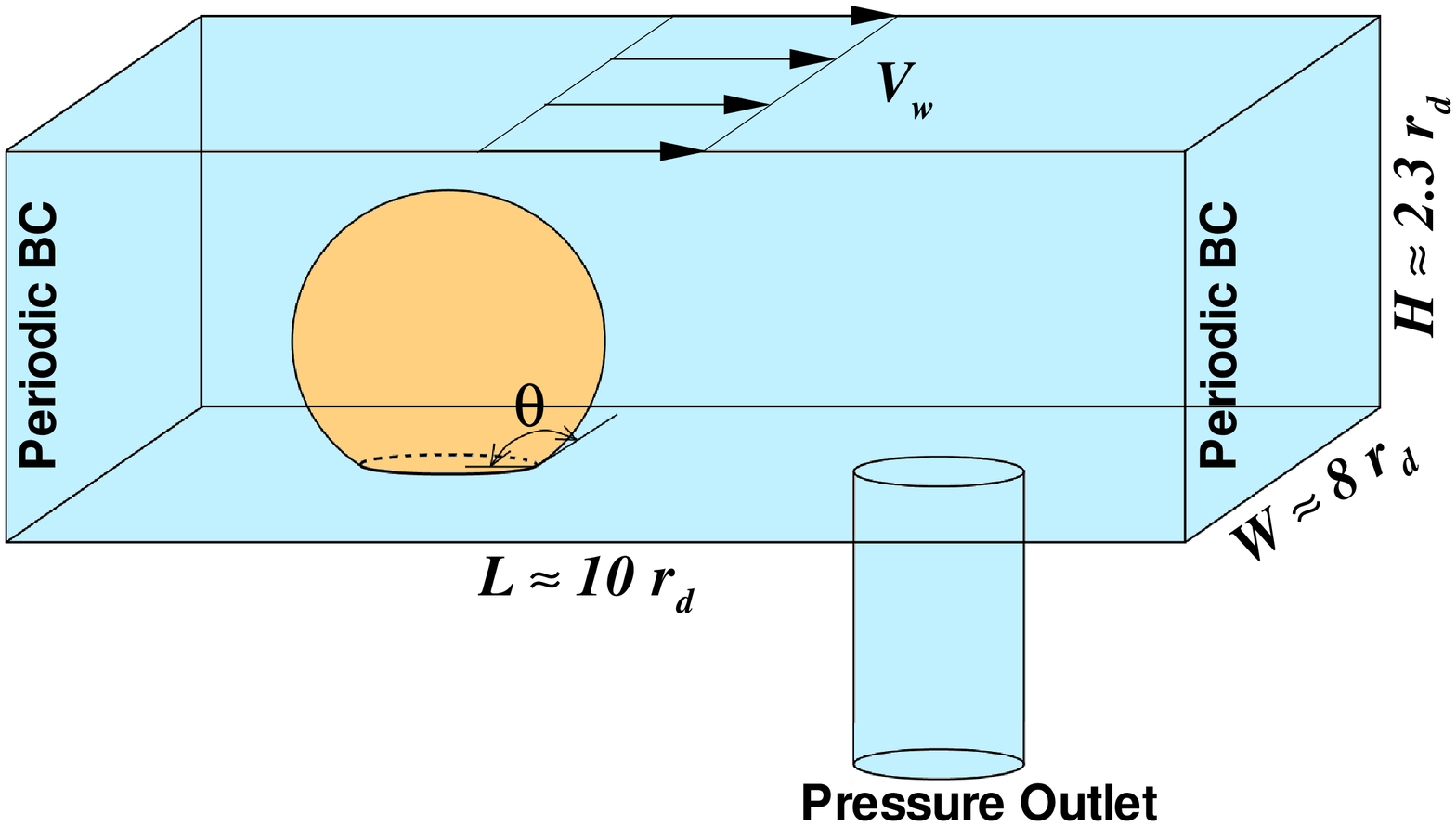}
\caption{(Color online) Schematic representation of the oil droplet
in the channel with the circular pore.  The shear flow is induced by
the upper wall moving with a constant velocity parallel to the
stationary lower wall.  The droplet radius is
$r_{d}=0.9\,\mu\text{m}$ and the pore radius is
$r_{p}=0.2\,\mu\text{m}$.  Periodic boundary conditions are applied
at the inlet and outlet, while a constant pressure is maintained at
the side walls. } \label{fig:channel_schematic}
\end{figure}

We next examine the combined effect of the transmembrane pressure
and crossflow velocity on the entry dynamics of an oil droplet into
a circular pore.  The computational setup is illustrated
schematically in Fig.\,\ref{fig:channel_schematic}.  The shear flow
is induced by translating the upper wall with a constant velocity.
In our simulations, the effective shear rate $\dot{\gamma}$ is
defined as the ratio of the upper wall velocity to the channel
height.  The relevant dimensionless numbers, the capillary and
Reynolds numbers, are estimated to be
$Ca=\mu_{w}\dot{\gamma}r_{d}/\sigma\lesssim0.03$ and
$Re=\rho_{w}\dot{\gamma}r^{\,2}_{d}/\mu_{w}\lesssim0.5$.  The width
of the channel is chosen to be about $8$ times larger than the
droplet radius in order to minimize finite size effects in the
lateral direction. Initially, the droplet is released upstream to
insure that the flow reaches a steady state before the droplet
approaches the pore.  At the same time, the transmembrane pressure
is set to a prescribed value, and the simulation continues until the
droplet either reaches the outlet, penetrates into the pore, or
breaks up.

\begin{figure}[t]
\includegraphics[width=14.0cm,angle=0]{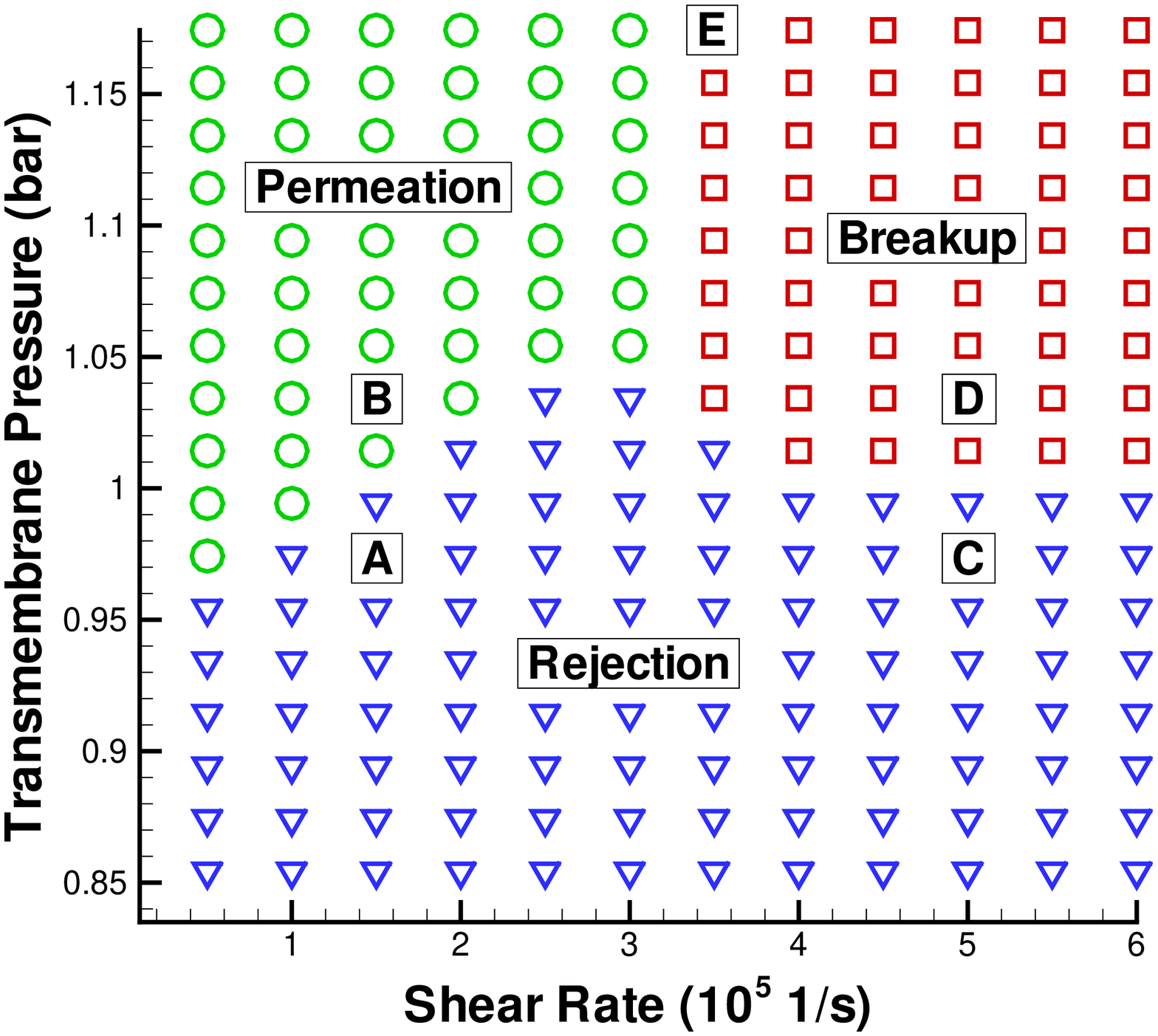}
\caption{(Color online)  The phase diagram of the transmembrane
pressure versus shear rate for the oil droplet with
$r_{d}=0.9\,\mu\text{m}$ and the circular pore with
$r_{p}=0.2\,\mu\text{m}$.  The contact angle is
$\theta=135^{\circ}$.  Each symbol represents a separate simulation
that corresponds to either $(\bigcirc)$ permeation, $(\square)$
breakup or $(\nabla)$ rejection.   Letters A, B, C, D, and E
indicate operating conditions for the series of snapshots shown in
Fig.\,\ref{fig:frames}.  } \label{fig:phase_diagram}
\end{figure}

\begin{figure}[t]
\includegraphics[width=16.0cm,angle=0]{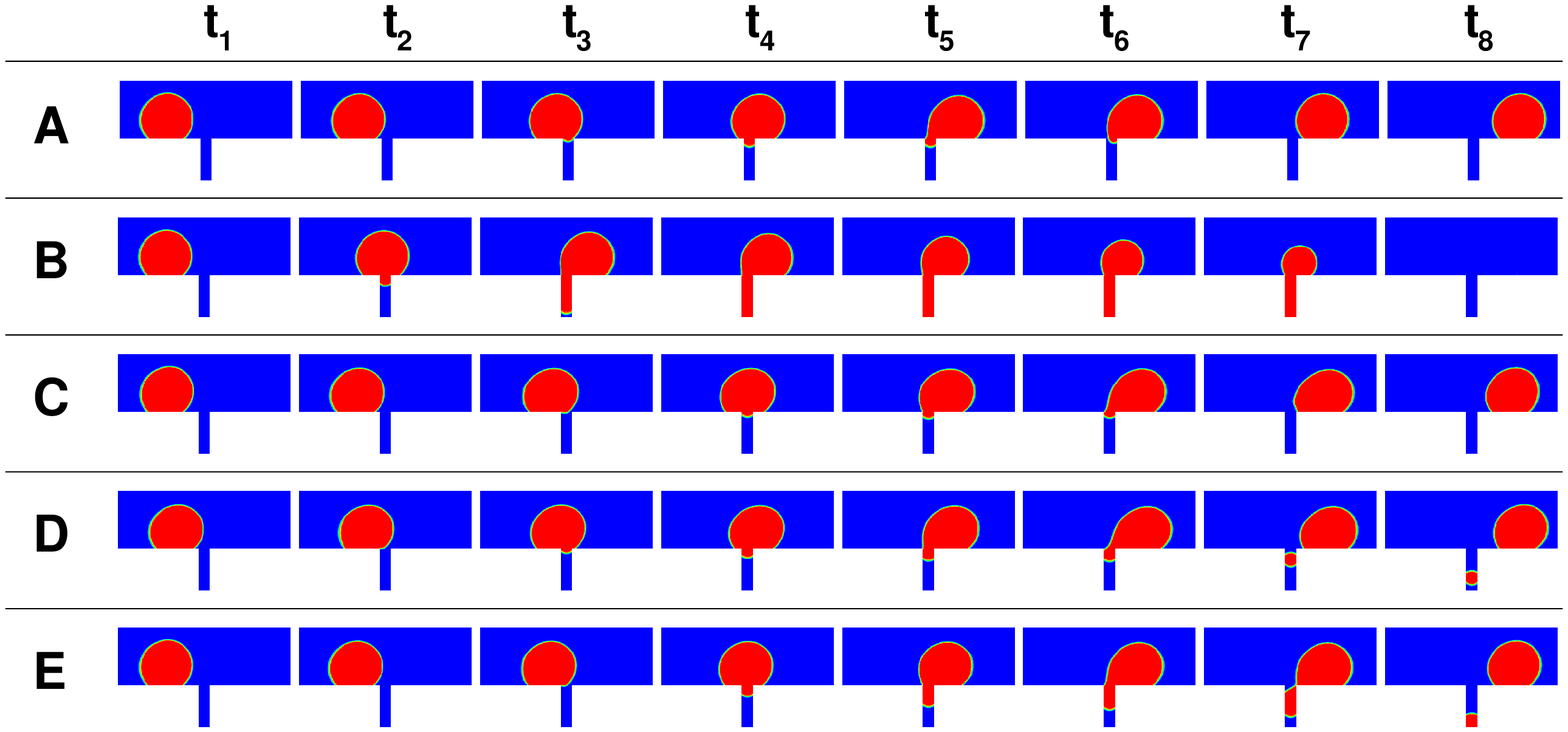}
\caption{ (Color online)  Sequences of snapshots of the sheared
droplet on the pressurized pore for different operating conditions
as indicated in Fig.\,\ref{fig:phase_diagram}.   The droplet radius
is $r_{d}=0.9\,\mu\text{m}$, the pore radius is
$r_{p}=0.2\,\mu\text{m}$, and the contact angle is
$\theta=135^{\circ}$.   The letters denote (A) rejection at low
shear rates (${\Delta}t\approx15\,{\mu}\text{s}$), (B) permeation
(${\Delta}t\approx15\,{\mu}\text{s}$), (C) rejection at high shear
rates (${\Delta}t\approx2\,{\mu}\text{s}$), (D) local breakup
(${\Delta}t\approx2.5\,{\mu}\text{s}$), and (E) breakup with necking
(${\Delta}t\approx4\,{\mu}\text{s}$).  } \label{fig:frames}
\end{figure}

The main results of this study are summarized in
Fig.\,\ref{fig:phase_diagram}, which shows the phase diagram for the
droplet rejection, permeation, and breakup depending on the
transmembrane pressure and shear rate.    The corresponding
snapshots of the droplet for five different cases (denoted by the
capital letters A, B, C, D, and E) are presented in
Fig.\,\ref{fig:frames}.    Below, we discuss the details of the
processes in the three different regions of the phase diagram and
provide an estimate of the leakage volume during the droplet
breakup.

In the permeation region shown in Fig.\,\ref{fig:phase_diagram}, the
transmembrane pressure is larger than the steamwise drag force, and
the oil droplet penetrates into the pore.   A series of snapshots at
point B in Fig.\,\ref{fig:frames} demonstrate the details of the
permeation process.   As observed in Fig.\,\ref{fig:frames}\,(B), at
first, the droplet partially penetrates into the pore and becomes
strongly deformed in the shear flow.   However, the droplet does not
breakup because its size above the pore decreases as the droplet
penetrates into the pore, and the viscous shear stress acts on a
progressively smaller surface area.

With increasing shear rate, the effect of viscous forces becomes
more important, resulting in strong deformation of the droplet shape
near the pore entrance.  We find that at sufficiently large
transmembrane pressures and $Ca\gtrsim0.015$, the oil droplet breaks
up.  In this case, the larger droplet is washed off downstream and
the smaller droplet enters the pore, and, as a result, the membrane
leaks.

Depending on the values of the transmembrane pressure and shear
rate, two different breakup regimes were observed.   The first
regime is bounded by the minimum breakup pressure, which is found
from Fig.\,\ref{fig:phase_diagram} to be $1.00\pm0.01\,\text{bar}$.
Above this pressure, a small fragment is detached and penetrates
into the pore while the main droplet is carried away by the shear
flow.   During this process, the droplet has a limited time to
deform, and, therefore, the breakup occurs only locally without
significant deformation, as shown in Fig.\,\ref{fig:frames}\,(D).

The breakup process at higher transmembrane pressures occurs in a
qualitatively different way [see Fig.\,\ref{fig:frames}\,(E)].   As
the droplet approaches the pore, it momentarily slows down and
remains at the pore entrance during the ``residence time".   In this
case, the effects of the drag and the transmembrane pressure are
relatively large and comparable with each other.   As a result, the
shape of the droplet is significantly deformed by the shear flow and
a thin bridge is formed between two parts of the stretched droplet.
This thinning is known as necking and it usually indicates the
initial stage of the breakup process~\cite{Xu08,Zwan09}.   For a
short time interval, the thin neck holds the two parts of the
droplet together. At the final stage of breakup, the neck gets
thinner and thinner near the the edge of the pore, and at some
point, it becomes unstable and the droplet breaks.

\begin{figure}[t]
\includegraphics[width=12.0cm,angle=0]{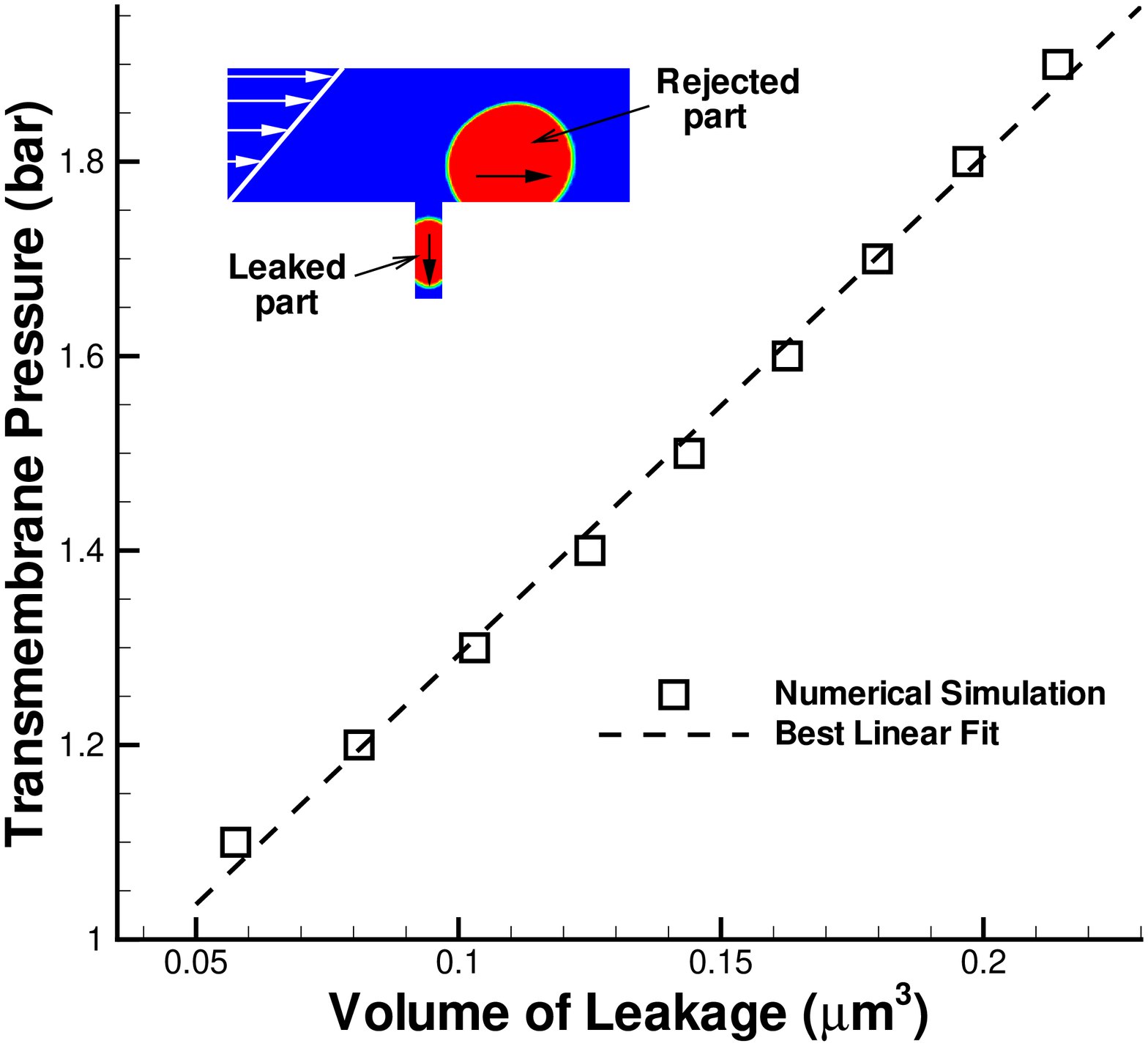}
\caption{(Color online) The leakage volume as a function of the
applied pressure for $r_{d}=0.9\,\mu\text{m}$,
$r_{p}=0.2\,\mu\text{m}$, and
$\dot{\gamma}=5\times10^{5}\,\text{s}^{-1}$.  The square symbols
indicate the numerical results and the dashed line is the best fit
to the data.  The inset shows a snapshot of the process shortly
after breakup of the droplet.  The contact angle of the oil droplet
in water is $\theta=135^{\circ}$. } \label{fig:tmp_vol}
\end{figure}

Visual inspection of the snapshots of the droplet near the pore
entrance revealed that, at a given shear rate in the breakup regime
in Fig.\,\ref{fig:phase_diagram}, the residence time is roughly
independent of the transmembrane pressure. Therefore, it is expected
that during the breakup process, the volume of leaked droplets is
proportional to the applied pressure. Figure\,\ref{fig:tmp_vol}
shows the leakage volume as a function of the transmembrane pressure
when $\dot{\gamma}=5\times10^{5}\,\text{s}^{-1}$.  Indeed, the
leakage volume is almost linearly proportional to the applied
pressure, indicating that the flow inside the pore is described by
the Hagen-Poiseuille equation.


The lower part of the phase diagram in Fig.\,\ref{fig:phase_diagram}
indicates operating conditions when the oil droplet is rejected by
the membrane and washed off by the shear flow.   An example of the
rejection process at low shear rates is presented in
Fig.\,\ref{fig:frames}\,(A).   Although the droplet partially
penetrates into the pore, the flow generates a force on the droplet
surface, which pulls the droplet away from the pore, resulting in
the droplet rejection.   As shown in Fig.\,\ref{fig:frames}\,(C),
the residence time at higher shear rates is reduced, and the droplet
is carried away by the flow without penetrating into the pore.

The threshold of permeation is determined by the competition between
the drag force and the transmembrane pressure.   Naturally, with
increasing shear rate, the drag force increases; and, therefore, it
is not surprising that the boundary curve separating the permeation
and rejection regions in Fig.\,\ref{fig:phase_diagram} increases
with shear rate.    However, it is difficult to estimate the exact
dependence of the drag force on the droplet because its shape
becomes strongly deformed in the shear flow.   We also comment that,
in the range of shear rates reported in
Fig.\,\ref{fig:phase_diagram}, the lift force is about an order of
magnitude smaller than the drag force~\cite{Xu05}.

The critical shear rate that marks the boundary between the
permeation and breakup regions in Fig.\,\ref{fig:phase_diagram} can
be estimated using simple force balance arguments.   In the absence
of gravity, the sheared droplet is subject to forces of surface
tension, Laplace pressure, drag, and lift.   Neglecting the lift
force and following the analysis in Ref.\,\cite{Xu05}, the torque
balance around the edge of the pore for the droplet configuration
depicted in Fig.\,\ref{fig:frames}\,(E\,$t_{6}$) can be written as
follows:
\begin{equation}
F_{D}\,d_{d}\,+\,(P_{2}-P_{1})\,A_{n}\,d_{n}-F_{\sigma}\,d_{n}\,=\,0,
\label{eq:torque_balance}
\end{equation}
where $d_{n}$ and $A_{n}=\pi\,d_{n}^{\,2}/4\,$ are the diameter and
cross-sectional area of the thinnest stable neck,
$F_{\sigma}=\pi\,\sigma\,d_{n}$ is the surface tension force around
the perimeter of the neck, and $P_{2}$ and $P_{1}$ are the pressures
inside the droplet and in the channel, respectively.   The Stokes
drag force on the spherical droplet,
$F_{D}\approx1.1\,\pi\,\mu_{w}\,\dot{\gamma}d_{d}^{\,2}$, is
estimated for the viscosity ratio $2.4$ and the average flow
velocity $\dot{\gamma}d_{d}/2$~\cite{Husny06}.  In our simulations,
the typical diameter of the thinnest stable neck is
$d_{n}\approx\,0.9\,d_{p}$  [see
Fig.\,\ref{fig:frames}\,(E\,$t_{6}$)].    Using
$d_{d}=1.7\,\mu\text{m}$, $\sigma=0.0191\,\text{N/m}$, and
$\mu_{w}=10^{-3}\,\text{kg/m\,s}$ in Eq.\,(\ref{eq:torque_balance}),
the critical shear rate is roughly estimated to be
$\dot{\gamma}\,\approx\,3.4\times10^{5}\,\text{s}^{-1}$, which is in
good agreement with the value
$\dot{\gamma}\,\approx\,3.2\times10^{5}\,\text{s}^{-1}$ obtained
numerically in Fig.\,\ref{fig:phase_diagram}.

The permeation, rejection, and breakup regions identified in the
phase diagram in Fig.\,\ref{fig:phase_diagram} can be useful for the
optimal design and operation of crossflow microfiltration systems.
It is apparent that the permeation region should be avoided for
filtration purposes.    The optimal performance of the
microfiltration system with maximum rejection is achieved in the
upper part of the rejection region where the large transmembrane
pressure results in high flux of water while the oil phase is
completely rejected.   However, the separation efficiency can be
increased at higher transmembrane pressures in the breakup region,
where the higher flux of water is accompanied by some oil leakage.

\section{Conclusions}

In this paper, the numerical simulations were carried out to
investigate the influence of the transmembrane pressure and
crossflow velocity on the entry dynamics of thin oil films and oil
droplets into pores of different cross-section.    We considered
hydrophilic membrane surfaces with contact angles of oil in water
greater than $90^{\circ}$.    The numerical method was validated
against the analytical solution for the critical pressure of
permeation of an oil droplet into a circular pore in the absence of
crossflow.   Furthermore, we found that the results of numerical
simulations of thin oil films on elliptical or rectangular pores
agree well with the theoretical prediction for the critical pressure
expressed in terms of geometric parameters of the pore
cross-section.   Also, examples of curved oil-water interfaces
inside elliptical and rectangular pores were discussed for different
aspect ratios and contact angles.

In the presence of crossflow above the membrane surface, we have
determined numerically the phase diagram for the droplet rejection,
permeation, and breakup as a function of the transmembrane pressure
and shear rate.    A detailed analysis of the droplet dynamics near
the pore entrance was performed in the three different regions of
the phase diagram.   We found that in the permeation region, the
transmembrane pressure is larger than the steamwise drag, and the
oil droplet penetrates into the circular pore.   With increasing
crossflow velocity, the shape of the droplet becomes strongly
deformed near the pore entrance; and, at sufficiently high
transmembrane pressures and shear rates, the droplet breaks up into
two fragments, one of which penetrates into the pore.   It was also
shown that during the breakup process, the residence time of the oil
droplet at the pore entrance is roughly independent of the
transmembrane pressure, and the volume of the leaked fragment is
nearly proportional to the applied pressure.    Finally, the
numerical value of the critical shear rate that separates the
permeation and breakup regions is in good agreement with an estimate
based on the force balance arguments.

\section*{Acknowledgments}

Financial support from the Michigan State University Foundation
Strategic Partnership Grant 71-1624 is gratefully acknowledged.  The
authors would like to thank V.~V. Tarabara for introducing the
problem and insightful discussions.   Computational work in support
of this research was performed at Michigan State University's High
Performance Computing Facility.

\bibliographystyle{prsty}

\end{document}